\documentclass[a4paper]{article}
\usepackage{amsmath,amssymb}

\begin{document}

\title{A new approach to modifying theories of gravity}

\author{Christian G.~B\"ohmer\footnote{c.boehmer@ucl.ac.uk} and Nicola Tamanini\footnote{n.tamanini.11@ucl.ac.uk}\\
        Department of Mathematics, University College London\\
        Gower Street, London, WC1E 6BT, UK}

\date{\today}

\maketitle

\begin{abstract}
We propose a new point of view for interpreting Newton's and Einstein's theories of gravity. By taking inspiration from Continuum Mechanics and its treatment of anisotropies, we formulate new gravitational actions for modified theories of gravity. These models are simple and natural generalisations with many interesting properties. Above all, their precise form can, in principle, be determined experimentally.
\end{abstract}

\section{Introduction}

Since Newton's formulation of the law of universal gravitation in the Principia 1687, the theory has been unchanged. Very few attempts have been made to modify Newtonian gravity, modified Newtonian dynamics being probably the one exception~\cite{Milgrom:1983ca}. It took almost 330 years until a generalisation of Newton's theory was successfully constructed. Einstein's theory of gravity was radically different to any previous physical theory. It abandoned the absolute notion of time and replaced the concept of force by the curvature of a four dimensional space. However, it took only a few years when the first modifications and extensions of Einstein's theory appeared in the literature, see~\cite{Goenner:2004se} for an excellent historical review. Ever since, modified theories of gravity have enjoyed a prominent role in theoretical physics. It is the aim of the current work to propose modifications of Newton's and Einstein's theories of gravity by applying the same idea to both of them.

Let us start with the actions of Newtonian gravity and its relativistic analogue, the Einstein-Hilbert action
\begin{align}
  S_{\rm Newton} &= \int \Bigl[ -\rho\varphi + \frac{1}{8\pi G} \delta^{ij}\partial_i\varphi\partial_j\varphi \Bigr] d^3x \,,
  \label{newton}\\
  S_{\rm EH} &= \int \Bigl[ \mathcal{L}_{\rm matter} + \frac{c^4}{16\pi G} g^{\mu\nu} R_{\mu\nu} \Bigr] \sqrt{-g} \, d^4x \,.
  \label{eh}
\end{align}
Variation of the Newtonian action with respect to the gravitation potential $\varphi$ yields the well-known Poisson equation $\Delta\varphi=4\pi G\rho$ where $\rho$ is a given matter distribution. Likewise, variations of the Einstein-Hilbert action with respect to the metric (which contains the gravitational potentials) gives the famous Einstein field equations $G_{\alpha\beta} = 8\pi G/c^4\, T_{\alpha\beta}$.

When comparing the Einstein-Hilbert action~(\ref{eh}) with other models in physics, it appears to be somewhat unnatural as one generally considers potential energies quadratic in the field strength, Hooke's law probably being the best known. However, when looking at the Einstein-Hilbert action, there are various very good reason for its form. Some of these can be motivated mathematically while others are simply observational. The field equations derived from it are close to Newton's theory of gravity and where the solutions deviate they are doing so in precisely the way to be in agreement with observations. However, there are observational facts which strongly indicate that our understanding of the gravitational force is far from complete. Dark matter and dark energy are two unknown forms of matter which are required to make the Universe work.

Looking at~(\ref{newton}) and~(\ref{eh}) we note that both actions contain a contraction with respect to the metric. In the Newtonian model we simply have the flat metric for Cartesian 3-space. We are now taking some inspiration from Continuum Mechanics and in particular constitutive equations which specify the various different models studied in this field. In Maxwell's electromagnetism the constitutive equations for instance define the form of the dielectric tensor $\mathbf{D} = \varepsilon \mathbf{E}$ where $\varepsilon$ is a rank 2 tensor in general and we should really write $D_i = \varepsilon_i{}^j E_j$. When working with the Faraday tensor and its corresponding excitation, one writes $H_{ij} = 1/4\, \epsilon_{ijmn} \chi^{mnkl} F_{kl}$ where for Maxwell's theory in vacuum $\chi^{mnkl} = \sqrt{-g}(g^{mk}g^{nl}-g^{nk}g^{ml})$, see~\cite{Hehl:2003}. In many simple applications $\varepsilon_i{}^j$ is taken to be proportional to the Kronecker delta which corresponds to a simple isotropic medium with dielectric constant $\epsilon$, namely $\varepsilon_i{}^j = \epsilon \delta_i^j$. Sometimes $\epsilon$ is allowed to vary throughout the medium. In other words, a rank 2 isotropic tensor is proportional to the metric tensor. We are using this observation to argue that the Newton and Einstein-Hilbert actions are based on the assumption of an isotropic medium. Since Nature has a strong tendency to be anisotropic, this appears to be rather unnatural. By doing so, we are changing the interpretation of the terms $\delta^{ij}$ and $g^{\mu\nu}$ in~(\ref{newton}) and~(\ref{eh}), respectively. We now view them much like the material tensors specified by constitutive equations which define the properties of the material. It is an interesting historical fact that F.~Klein noted in a letter to Hilbert, as early as 1917, that the Ricci scalar in the Einstein-Hilbert action can be written as $\chi^{mnkl} R_{mnkl}$, see~\cite{Klein:1917}. For a recent paper inspired by similar thoughts, see~\cite{Battye}.

Our approach follows the principal ideas of Brans and Dicke~\cite{Brans:1961sx}. They suggested a model where the gravitational coupling was allowed to vary in space and time and it was viewed as an additional dynamical degree of freedom in the theory. The original Brans-Dicke model can be viewed as the isotropic limit of out model, we simply choose $C^{\mu\nu} = \phi(x^\alpha)\, g^{\mu\nu}$ with the main difference that we do not treat the new degrees of freedom as dynamical.  

We might want to, for sake of concreteness, speak of the properties of the vacuum here. However, we would like to be very careful and point out that this change of viewpoint has a variety of philosophical implications when interpreting the modified theories. We would like to keep these issues aside for now and proceed with the formulation of the theories. The assumption of an isotropic vacuum is certainly well supported by a host of experimental observations. However, there is clearly room for some improvement, in particular in general relativity. Many of the well-known modifications are is some sense rather severe. Field equations become higher than second order, new fields are introduced, locality is broken, local Lorentz invariance is broken etc. We will show that our theories are completely harmless and retain all desired properties.

\section{Modified gravity using continuum mechanics}

One of the most conservative modifications of Newton's and Einstein's theories we can think of is therefore the following
\begin{align}
  S_{\rm Newton} &= \int \Bigl[ -\rho\varphi + \frac{1}{8\pi G} C^{ij}\partial_i\varphi\partial_j\varphi \Bigr] d^3x \,,
  \label{newton1}\\
  S_{\rm EH} &= \int \Bigl[ \mathcal{L}_{\rm matter} + \frac{c^4}{16\pi G} C^{\mu\nu} R_{\mu\nu} \Bigr] \sqrt{-g} \, d^4x \,,
  \label{eh1}
\end{align}
where $C^{ij}$ and $C^{\mu\nu}$ are symmetric rank 2 tensors which contain information about the underlying structure of the theory. In Continuum Mechanics such objects are often referred to as material tensors or elastic coefficients~\cite{Green:1963}. We will stick with this well established notation and should clearly distinguish it from the energy-momentum tensor $T_{\alpha\beta}$. We should emphasise that in this approach to gravity we are able to assume any symmetry for the metric, for instance spherical symmetry or homogeneity and isotropy. However, the symmetry of the metric is independent of the symmetry of the elastic coefficients.

Let us firstly consider variations of the modified Newtonian theory~(\ref{newton1}) with respect to the potential $\phi$. We find
\begin{align}
  \partial_i \bigl(C^{ij} \partial_j \varphi \bigr) = 4\pi G \rho\,,
  \label{field1}
\end{align}
which reduces to the Poisson equation if we choose $C^{ij} = \delta^{ij}$. As $C^{ij}$ can in principle be an arbitrary tensor, solutions to this equation may be quite different. As a simple example, let us consider the case when $C^{ij} = \mbox{diag}(c_1,c_2,c_3)$ with the $c_i$ being constants. In continuum mechanics or solid state physics one would speak of a crystal with three principal propagation directions. In this case, the field equation~(\ref{field1}) can be reduced to the Poisson equation by rescaling the coordinates. Looking for a radially symmetric solution gives the interesting result
\begin{align}
  \varphi \propto \frac{1}{\sqrt{c_1 x^2 + c_2 y^2 + c_3 z^2}}\,.
\end{align}
In this case the gravitational field of a massive body is ellipsoidal instead of being spherical. However, the strength of this effect depends on the values of the constants. If for instance, the numerical values of the three constants $c_1,c_2,c_3$ are very close to each other, then the gravitational field will look spherical unless very large distances are taking into account.

It should also be noted that the components of $C^{ij}$ do not have to be constants. They can be functions of the coordinates. Thus, the form of the gravitational field may be different in different regions of space, however, the gravitational law~(\ref{field1}) would still be universal. If we consider $C^{ij} = \chi(t,x,y,z) \delta^{ij}$, the field equations are
\begin{align}
  \Delta \varphi = \frac{4\pi G}{\chi}\rho - \frac{\partial^i\varphi\,\partial_i\chi}{\chi}\,.
\end{align}
If the function $\chi$ is slowly varying in space $|\partial_i\chi| \ll 1$, we would have a theory which would be in very good agreement with Newton's theory on smaller scales, like the solar system. Note also that assuming a time varying $\chi$ would correspond to a time dependent gravitational constant, compare with Dirac's large number hypothesis~\cite{Dirac:1938mt}.

Let us now consider the action~(\ref{eh1}). We assume that the tensor $C^{\mu\nu}$ may depend on the metric, this is important so that the limit of general relativity can be recovered. Note that the variation of the Ricci tensor does not give a surface term anymore because $C^{\mu\nu}$ is not the metric. One has to integrate by parts twice (the Ricci tensor contains second derivatives of the metric) and move the derivatives action on the metric variations to act on the matter tensor. Denoting
\begin{align}
  \Sigma^{\mu\nu\alpha\beta} = -\frac{\delta C^{\mu\nu}}{\delta g_{\alpha\beta}} \,,
  \label{eqn2}
\end{align}
the whole calculation yields the field equation
\begin{multline}
  \Sigma^{\mu\nu\alpha\beta} R_{\mu\nu} - \frac{1}{2}g^{\alpha\beta} C^{\mu\nu} R_{\mu\nu} + \frac{1}{2} \Box C^{\alpha\beta} \\+ \frac{1}{2}g^{\alpha\beta} \nabla_{\mu} \nabla_{\nu} C^{\mu\nu}
  - \nabla_{\mu} \nabla^{(\alpha} C^{\beta)\mu} = \frac{8\pi G}{c^4} T^{\alpha\beta} \,.
  \label{eqn3}
\end{multline}
where we have also included the energy-momentum tensor of matter.

Recall that in General Relativity the field equations imply the conservation equations by virtue of the twice contracted Bianchi identities. This does no longer hold due to the presence of the material tensor $C^{\mu\nu}$ in the field equations. Nonetheless, we can take the covariant derivative of the field equation~(\ref{eqn3}) with respect to $\nabla_{\alpha}$ and find the following conservation equation
\begin{multline}
  \nabla_{\alpha} \Sigma^{\mu\nu\alpha\beta} R_{\mu\nu}
  +\Sigma^{\mu\nu\alpha\beta} \nabla_{\alpha} R_{\mu\nu}
  -\frac{1}{2} g^{\alpha\beta} R_{\mu\nu} \nabla_{\alpha} C^{\mu\nu} \\
  -C^{\alpha\sigma} \nabla_{\alpha} R^{\beta}{}_{\sigma}
  -R^{\beta}{}_{\sigma} \nabla_{\alpha} C^{\alpha\sigma} 
  = \frac{8\pi G}{c^4} \nabla_{\alpha} T^{\alpha\beta} \,.
  \label{eqn4}
\end{multline}
This implies that the conservation equations for $T^{\alpha\beta}$ is no longer a direct consequence of the gravitational field equations alone. However, if we assume that the energy-momentum tensor is derived from a diffeomorphism invariant matter action, then the energy-momentum tensor is covariantly conserved in view of Noether's theorem 
\begin{align}
  \nabla_{\alpha} T^{\alpha\beta} = 0 \,.
  \label{cons}
\end{align}
We will assume this henceforth. Therefore, the conservation equation~(\ref{cons}) imposes constraints on the components of $C^{\mu\nu}$ via~(\ref{eqn4}) and thus its components cannot be chosen completely arbitrarily. For concreteness, we define
\begin{multline}
  J^{\beta} = \nabla_{\alpha} \Sigma^{\mu\nu\alpha\beta} R_{\mu\nu}
  +\Sigma^{\mu\nu\alpha\beta} \nabla_{\alpha} R_{\mu\nu}
  -\frac{1}{2} g^{\alpha\beta} R_{\mu\nu} \nabla_{\alpha} C^{\mu\nu} \\
  -C^{\alpha\sigma} \nabla_{\alpha} R^{\beta}{}_{\sigma}
  -R^{\beta}{}_{\sigma} \nabla_{\alpha} C^{\alpha\sigma} \,,
  \label{eqnj}
\end{multline}
which gives the additional consistency equation
\begin{align}
  J^{\beta} = 0 \,.
  \label{eqnj2}
\end{align}

It should also be emphasised that one has to be careful when choosing the tensor $C^{\mu\nu}$ and the derived quantity $\Sigma^{\mu\nu\alpha\beta}$. Namely, there is a conceptual difference between prescribing the tensor $C^{\mu}_{\nu}$ from prescribing $C^{\mu\nu}$. In the former case, we find that $C^{\mu\nu} = g^{\mu\sigma} C^{\nu}_{\sigma}$ and thus $C^{\mu\nu}$ has an explicit dependence on the metric which in turn will yield a non-trivial form for $\Sigma^{\mu\nu\alpha\beta}$. On the other, if we specify $C^{\mu\nu}$ {\it a priori}, then this tensor does not depend on the metric and hence the resulting variational derivative would be zero. One has to carefully distinguish both cases when analysing the field equations.

It is very difficult to analyse the field equations~(\ref{eqn3}) given general $\Sigma^{\mu\nu\alpha\beta}$ and $C^{\mu\nu}$. It is also not clear what would constitute a good choice for those quantities. As a first attempt to understand the theory, we assume $C^{\mu\nu}$ to be conformally related to the inverse metric
\begin{align}
  C^{\mu\nu} = \phi(x^\alpha)\, g^{\mu\nu} \,,
  \label{eqn4a}
\end{align}
where $\phi(x^\alpha)$ is a scalar function depending on the coordinates. This choice simplifies the field equation considerably and they are now given by
\begin{align} 
  \phi\, G^{\alpha\beta} +g^{\alpha\beta} \Box\phi - \nabla^{\alpha} \nabla^{\beta} \phi
  = \frac{8\pi G}{c^4} T^{\alpha\beta} \,.
  \label{eqn5}
\end{align}
Upon division by the field $\phi$ and solving for the Einstein tensor, the resulting equation 
\begin{align} 
  G^{\alpha\beta}  = \phi^{-1}\, \frac{8\pi G}{c^4} T^{\alpha\beta} + 
  \phi^{-1}\, \bigl(\nabla^{\alpha} \nabla^{\beta} \phi - g^{\alpha\beta} \Box\phi\bigr)\,.
  \label{eqn6}
\end{align}
shows similarities with non-minimally coupled scalar field theories~\cite[see Eq.~(11)]{Brans:1961sx}. The main difference between the current theory and most other approaches is that $\phi$ is not a dynamical degree of freedom because $C^{\mu\nu}$ is a prescribed tensor and has not corresponding equations of motion. The conformal model appears to agree with~\cite{Brans:1961sx} in the limit when the Brans-Dicke parameter $\omega \rightarrow 0$. However, as variations with respect to $\phi$ are not considered, there is no propagation equation for the scalar field. By substituting~(\ref{eqn4a}) back into the action~(\ref{eh1}), we note that this simply corresponds to assuming the gravitational constant to be varying in time and space, see for instance~\cite{Dirac:1938mt}. Varying constants models are generally based on non-minimally coupled scalar field with kinetic term similar to Brans-Dicke theory, see~\cite{Barrow:1996kc}. 

\section{A Schwarzschild like solution}

Let us start by considering a static and spherically symmetric vacuum spacetime described by the metric
\begin{align}
  ds^2 = -e^{\nu(r)} dt^2 + e^{\mu(r)} dr^2 + r^2 d\Omega^2 \,,
\end{align}
where $d\Omega^2$ is the usual line element of the two-sphere. We also assume $\phi = e^{\xi(r)}$. When the analogue situation is analysed in General Relativity where $\xi \equiv 0$, one finds two independent equations which determine the two unknown functions $\nu(r)$ and $\mu(r)$. In this model, there is the additional degree of freedom $\xi$ and fortunately, there are now three independent equations. These are given by
\begin{align}
  -\frac{e^{\mu}}{r^2} + \frac{1}{r^2} - \frac{1}{2} \mu' \xi' - \frac{\mu'}{r} + \xi'' + (\xi')^2 + 2\frac{\xi'}{r} &= 0 \,,
  \label{sph1}\\
  -\frac{e^{\mu}}{r^2} + \frac{1}{r^2} + \frac{1}{2} \nu' \xi' + \frac{\nu'}{r} + 2\frac{\xi'}{r} &= 0 \,,
  \label{sph2}\\
  -\frac{1}{4} \mu' \nu ' - \frac{1}{2} \mu' \xi' - \frac{\mu'}{2 r} + \frac{\nu''}{2} + \frac{1}{2}\nu' \xi' + \frac{1}{4} (\nu')^2 & \nonumber \\+ \frac{\nu'}{2 r} + \xi'' + (\xi')^2 + \frac{\xi'}{r} &= 0 \,.
  \label{sph3}
\end{align}
Eqn.~(\ref{sph2}) can be solved for the function $\mu$ which can then be substituted into the other two equations. Combining those linearly gives the condition $\nu' \propto \xi'$ which then allows us to reduce this problem to a single differential equation. While separation of variable and subsequent integration is possible, the resulting equation cannot be solved analytically for the unknown functions due to its high nonlinearity. However, by assuming $\xi' = -C^2 \nu'$ with $C \ll 1$ we can find an approximate solution to the field equations which is given by
\begin{align}
  e^{\nu} &= 1 - \frac{C}{r} + O(C^4)\,,\\
  e^{-\mu} &= 1 - \frac{C}{r} + \frac{2C^3}{r} + O(C^4)\,,\\
  e^{\xi} &= 1 + \frac{C^3}{r} + O(C^4)\,.
\end{align}
One easily verifies that this solution satisfies the field equations~(\ref{sph1})--(\ref{sph3}) up to $O(C^4)$. By choosing $C = 2 G M$ we arrive at a Schwarzschild like solution with only a small difference. Clearly, this difference of the order of $C^3$ and therefore it would be very difficult to distinguish between this metric and the Schwarzschild metric using solar system tests. This is a promising result which indicates that this theory can pass solar system tests without great difficulty.

\section{Cosmology}

\subsection{Conformal model}

Next, we want to study the cosmological implications of field equations~(\ref{eqn3}). Similar to the above we assume $C^{\mu\nu} = \phi(t)\, g^{\mu\nu}$ and consider a FLRW universe
\begin{align}
  ds^2 = -dt^2 + a^2(t)\, \frac{dx^2+dy^2+dz^2}{\bigl(1+\frac{k}{4}r^2\bigr)^2} \,,
\end{align}
where $a(t)$ is the scale factor and $r^2 = x^2 + y^2 + z^2$. This yields the following cosmological field equations
\begin{align}
  3H^2\phi+3H\dot\phi+3\frac{k\phi}{a^2} &= \frac{8\pi G}{c^4} \rho \,, 
  \label{eqn001} \\
  -\ddot\phi-2\phi\frac{\ddot a}{a} -2H\dot\phi-H^2\phi -\frac{k\phi}{a^2} &= \frac{8\pi G}{c^4} p \,, 
  \label{eqn002}
\end{align}
where an overdot denotes differentiation with respect to time derivative and $H$ is the Hubble parameter $H=\dot a/a$. Moreover, the conservation equation~(\ref{cons}) gives
\begin{align}
  \dot\rho + 3H (\rho+p) = 0 \,,
  \label{eqn013}
\end{align}
while the consistency equation~(\ref{eqnj2}) for $J^0$ becomes
\begin{align}
  3 \dot{\phi} \Bigl(\dot{H} + 2 H^2 + \frac{k}{a^2}\Bigr) = 0 \,.
\end{align}
The other three components vanish identically $J^1 = J^2 = J^3 = 0$.

Irrespective of the choice of matter in this model, the consistency equation~(\ref{eqnj2}) either implies that $\dot{\phi} = 0$ which is equivalent to General Relativity, or 
\begin{align}
  \dot{H} + 2 H^2 + \frac{k}{a^2} = 0 \,.
\end{align}
The solutions to this differential equation are given by
\begin{alignat}{2}
  a(t) &= a_0 \sqrt{t-t_0} &\qquad k &= 0 \,,
  \label{scale0}\\
  a(t) &= \sqrt{a_0^2 - k(t-t_0)^2} &\qquad k &= \pm 1 \,,
\end{alignat}
and correspond to a radiation dominated universe. Next, we have to check whether these solutions are consistent with the remaining equations~(\ref{eqn001}), (\ref{eqn002}) and~(\ref{eqn013}). Firstly, we start looking for vacuum solutions, $\rho = p =0$. The conservation equation~(\ref{eqn013}) is trivially satisfied and one verifies that 
\begin{alignat}{2}
  \phi(t) &= \frac{\phi_0}{\sqrt{(t-t_0)}} &\qquad k &= 0 \,,\\
  \phi(t) &= \frac{\phi_0 (t-t_0)}{\sqrt{a_0^2 - k(t-t_0)^2}} &\qquad k &= \pm 1 \,,
\end{alignat}
is a solution to the remaining field equations.

Under the assumption that $C^{\mu\nu} = \phi(t)\, g^{\mu\nu}$ and $\rho = p =0$ we cannot find an accelerating solution to the field equations. We could, of course, add regular matter to the field equations and seek other solutions. Let us briefly consider the situation where we include an incompressible perfect fluid or dust, for simplicity we consider $k=0$ only. The scale factor is unchanged and given by~(\ref{scale0}). The conservation equation~(\ref{eqn013}) implies the standard relation $\rho \propto 1/a^3$ and it turns out that 
\begin{align}
  \rho(t) = \frac{\rho_0}{(t-t_0)^{3/2}}\,, \qquad 
  \phi(t) = \frac{16\pi \rho_0 t + 3 \phi_0}{3\sqrt{t-t_0}} \,,
\end{align}  
is a solution to the field equations.

\subsection{Fluid like model}

However, we are particularly interested in the vacuum equations without ordinary matter. The reason for this is simply that we want to show that an additional non-dynamical structure in the theory suffices to get a dynamical universe. One can easily find such solutions by introducing an extra degree of freedom in the material tensor. Let us choose the elastic coefficients to be
\begin{align}
  C^{\mu}_{\nu} = \mbox{diag}(-\varrho,\sigma,\sigma,\sigma) \,,
  \label{c0}
\end{align}
which means that $C^{\mu\nu} = \frac{1}{2}(g^{\mu\sigma} C^{\nu}_{\sigma}+g^{\nu\sigma} C^{\mu}_{\sigma})$ and thus we have an explicit dependence on the metric, implying a non-zero $\Sigma^{\mu\nu\alpha\beta}$ given by 
\begin{align}
  \Sigma^{\mu\nu\alpha\beta} = \frac{1}{4} \bigl( 
  g^{\mu\alpha} C^{\nu\beta} + g^{\nu\alpha} C^{\mu\beta} + 
  g^{\mu\beta} C^{\nu\alpha} + g^{\nu\beta} C^{\mu\alpha} 
  \bigr)\,.
\end{align} 
One can interpret the quantities $\varrho$ and $\sigma$ as the energy density and pressure of the vacuum, thereby specifying its internal structure.

The cosmological field equations, the conservation equation and the consistency equation of this model are given by
\begin{align}
  -H^2 \varrho - \frac{1}{2} H (\dot{\varrho} - \dot{\sigma}) + 
  \frac{k \sigma}{a^2} &= \frac{8\pi G}{3c^4} \rho\,,
  \label{c1}\\
  3 H^2 \varrho + 2(\varrho H)\dot{} 
  +\frac{1}{2}(\ddot{\varrho} - \ddot{\sigma}) 
  - \frac{k \sigma}{a^2} &= \frac{8\pi G}{c^4} p \,,
  \label{c2}\\
  H^2 (\dot{\varrho} - 3\dot{\sigma}) + 
  \dot{H}(\dot{\varrho} - \dot{\sigma}) 
  - \frac{2 k \dot{\sigma}}{a^2} &= 0 \,,
  \label{c2a}\\
  \dot{\rho} + 3H(\rho + p) &= 0 \,,
  \label{c2b}
\end{align}
where the conservation equation can be derived from the first three equations. In the analogue situation in General Relativity we would have two independent equations for three unknown functions while for this model we have one additional equation and two additional degrees of freedom. 

To begin with, we consider the vacuum case $\rho = p =0$ which eliminates one equation and two degrees of freedom. Thus, we are left with two independent equations and three unknown functions. In order to close this system, we choose a linear equation of state $\sigma = w \varrho$ for the spacetime structure. 

Now, we can solve Eq.~(\ref{c1}) for $\dot{\varrho}$ and substitute this result into Eq.~(\ref{c2}) to arrive at a single differential equation in $a(t)$. One can also check that substitution of $\dot{\varrho}$ into Eq.~(\ref{c2a}) leads to the same differential equation, confirming that these equations are indeed not independent. This differential equation is given by
\begin{align}
  (\dot{a}^2 - k w)\Bigl((w-1)a\ddot{a} + 2w(k+\dot{a}^2)\Bigr) = 0\,,
  \label{diff}
\end{align}
where in the derivation we assumed $\varrho \neq 0$, $w \neq 1$ and $\dot{a} \neq 0$. This differential equation is the product of two equations and thus can be solved by finding a solution to either of the two equations. The first one is easily solved by
\begin{align}
  a(t) = \sqrt{kw}(t-t_0) \,, 
\end{align}
and is valid only if $kw>0$. 

The second differential equation in~(\ref{diff}) cannot be solved analytically for arbitrary $k$ and $w$ due to the non-linear nature of the equation. For $w=1/3$ for instance we can find the two solutions
\begin{align}
  a(t) &= \frac{1}{2}\Big(
  \frac{1}{\alpha^2} e^{\alpha(t-t_0)} + k e^{-\alpha(t-t_0)} \Bigr) \,,\\
   a(t) &= \frac{1}{2}\Big(
   k e^{\alpha(t-t_0)} + \frac{1}{\alpha^2} e^{-\alpha(t-t_0)} \Bigr) \,,
\end{align}
which are valid for all $k$. For $w=0$ one only finds one solution $a(t) = a_0 t$. For $k=0$ one can solve the differential equation for all $w$ and its solution is given by
\begin{align}
  a(t) = a_0 (t-t_0)^{\frac{w-1}{3w-1}} \,,
\end{align}
which is well defined provided that $w \neq 1/3$ and corresponds to a power-law solution. For $0 < w < 1/3$ this would correspond to an accelerated solution. The case $w=1/3$ needs to be treated separately. As the exponent becomes very large, one would expect this to correspond to exponential functions, and indeed in this case
\begin{align}
  a = a_0 e^{\lambda t}\,.
  \label{c7}
\end{align}
Thus, we were able to find solutions of the field equations modelling a universe which can undergo periods of accelerated expansion, without the need to introduce any forms of matter. All we have done is to add an additional non-dynamical structure to the theory on a very fundamental level.

\subsection{Kasner type model}

We are now considering a Kasner type metric given by
\begin{align}
  ds^2 = -dt^2 + t^{2p_1} dx^2 + t^{2p_2} dy^2 + t^{2p_3} dz^2\,,
  \label{ka1}
\end{align}
and assume the material tensor $C^{\mu}_{\nu}$ to be of the form
\begin{align}
  C^{\mu}_{\nu} = \mbox{diag}(-1,c_1(t),c_2(t),c_3(t)) \,.
  \label{ka2}
\end{align}
The field equations of this system are quite complicated. However, one notes that all field equations contain terms of the form $t c_{i}'(t)$ and $t^2 c_{i}''(t)$ which indicate that one can arrive at algebraic equations by choosing 
\begin{align}
  c_i(t) = 2 \gamma_i \log(t) \,,
  \label{ka3}
\end{align}
where the $\gamma_i$ are constants. With this additional assumption~(\ref{ka3}), and considering a vacuum $\rho = p = 0$, the field equations are given by
\begin{align}
  p_1 \gamma_1 + p_2 \gamma_2 + p_3 \gamma_3 
  - p_1 p_2 - p_1 p_3 - p_2 p_3 &= 0 \,,
  \label{kaA}\\
  (p_1 + p_2 + p_3 -1) (\gamma_1-\gamma_2+p_1-p_2) &= 0 \,,
  \label{kaB}\\
  (p_1 + p_2 + p_3 -1) (\gamma_1-\gamma_3+p_1-p_3) &= 0 \,,
  \label{kaC}\\
  (p_1 + p_2 + p_3 -1) (\gamma_2-\gamma_3+p_2-p_3) &= 0 \,,
  \label{kaD}\\
  (p_1 + p_2 + p_3 -1) (p_1 \gamma_1 + p_2 \gamma_2 + p_3 \gamma_3) &= 0 \,.
  \label{kaE}
\end{align}
The structure of these equations is quite interesting as 4 of the 5 equations can be solved immediately by assuming the Kasner condition $p_1 + p_2 + p_3 = 1$. Note that in General Relativity this condition is necessary to solve the field equations. The condition $p_1 + p_2 + p_3 = 1$ and the remaining equation~(\ref{kaA}) give two algebraic relation for the 6 free parameters. 

However, we can also find solutions to Eqs.~(\ref{kaA})--(\ref{kaE}) without the Kasner condition. We start by assuming that $p_1 + p_2 + p_3 \neq 1$ which allows us to divide Eqs.~(\ref{kaB})--(\ref{kaE}) by the factor $(p_1 + p_2 + p_3 -1)$. We note that the three Eqs.~(\ref{kaB})--(\ref{kaD}) are not independent as $(\ref{kaB}) - (\ref{kaC}) + (\ref{kaD}) = 0.$ Hence, we are left with four independent equations. 

One easily verifies that a solution can be written in the form
\begin{align}
  \gamma_1 = \frac{p_2^2}{p_1 + p_2}\,, \qquad \gamma_2 &= \frac{p_1^2}{p_1 + p_2}\,, \qquad
  \gamma_3 = p_1 + p_2\,, 
  \label{ka4}\\
  p_3 &= -\frac{p_1 p_2}{p_1 + p_2}\,.
  \label{ka5}
\end{align}
We should remark that Eq.~(\ref{ka5}) does imply that one of the three $p_i$ has to be negative. While the additional structure due to the material tensor $C^{\mu}_{\nu}$ changes the underlying conceptual physics substantially, the solution shows many similarities with General Relativity. 

\section{Conclusions}

It is tempting to argue that we introduced a form of matter through the back door by choosing our material tensor, the elastic coefficients. However, it is far from clear whether this is indeed the case. Note that $C^{\mu\nu}$ is not a dynamical variable and thus it cannot be interpreted simply as matter. In Continuum Mechanics and when working with crystal symmetries, the tensor $C^{\mu\nu}$ is said to encode the symmetry properties of the material, in our case the vacuum. We simply say that the vacuum as we know it may have an internal structure which is specified by $C^{\mu\nu}$. We are breaking away from the assumption that the vacuum is isotropic and structureless.

The form of the elastic coefficients can in principle be determined observationally. In the context of cosmology, one could start with the specific $C^{\mu\nu}$ given by~(\ref{c0}) and assume it to be close to the metric $g^{\mu\nu}$. When one considers models where deviations from General Relativity will vary with cosmological time, it would be most interesting to see how observational data would determine the form of $C^{\mu\nu}$ which provides the best fit to the data. Using our approach to gravity, we will be able to use observations directly to specify the model instead of guessing new theories and deriving their implications on a case by case basis. Ultimately, experiments and observations will determine the correct theory.

\subsection*{Acknowledgements}
We thank Friedrich Hehl and Dmitri Vassiliev for useful discussions. We would also like to thank the anonymous referees who have given us very valuable feedback.

\end{document}